\documentclass[aps,prl,twocolumn,superscriptaddress]{revtex4}

\usepackage[dvips]{graphicx}
\usepackage{color}

\begin{document}

\title{Observation of Spin Glass Dynamics in Dilute LiHo$_x$Y$_{1-x}$F$_4$}

\author{J. A. Quilliam}
\author{S. Meng}
\author{C. G. A. Mugford}
\author{J. B. Kycia}

\affiliation{Department of Physics and Astronomy and Guelph-Waterloo Physics Institute, University of
Waterloo, Waterloo, ON N2L 3G1 Canada}

\affiliation{Institute for Quantum Computing, University of
Waterloo, Waterloo, ON N2L 3G1 Canada}

\date{\today}

\begin{abstract}

AC susceptibility measurements are presented on the dilute, dipolar coupled, Ising magnet LiHo$_x$Y$_{1-x}$F$_4$ for a concentration $x = 0.045$.  The frequency and temperature dependences of the susceptibility show characteristic glassy relaxation.  The absorption spectrum is found to broaden with decreasing temperature suggesting that the material is behaving as a spin glass and not as an exotic spin liquid as was previously observed.  A dynamical scaling analysis suggests a spin glass transition temperature of 42.6 mK $\pm 2$ mK with an exponent $z\nu = 7.8 \pm 0.23$.

\end{abstract}


\maketitle


Spin glass behavior is an effect resulting from quenched disorder that has been extensively studied for many years and is largely well
understood.  It was therefore surprising that the system LiHo$_x$Y$_{1-x}$F$_4$, a seemingly ideal dilute, dipolar coupled Ising magnet, would exhibit an unusual spin liquid state at a low concentration of magnetic ions, $x = 0.045$~\cite{Reich1987, Reich1990}.  This spin liquid, or ``antiglass'' phase was most notably characterized by a narrowing of the absorption spectrum $\chi''(\omega)$ as temperature was lowered.  This was found in stark contrast with a higher concentration ($x = 0.167$) spin glass state where the absorption spectrum broadens~\cite{Reich1990,Wu1991,Wu1993} and with theoretical predictions that the spin glass state should persist to 0 concentration in such a system with long-range interactions~\cite{Aharony1981}.

Later publications found numerous interesting effects in the spin liquid state including very narrow absorption spectra (narrower than what can be accounted for with models of glassy relaxation) with strong asymmetry, ringing magnetization oscillations, sharp features in the specific heat and a $T^{-0.75}$ power law in the dc limit of the susceptibility~\cite{Ghosh2002,Ghosh2003}.

Recently there has been a large amount of activity on this series of materials, both theoretical and experimental.  In particular there is an ongoing debate on the existence of a spin glass transition at any value of $x$ in the LiHo$_x$Y$_{1-x}$F$_4$ series~\cite{Biltmo2007, Jonsson2007,AnconnaTorres2008,Jonsson2008}.  Much research has also taken place attempting to understand the effects of transverse magnetic field on the purported spin glass state (see for example~\cite{Wu1993,Schechter2005,Tabei2006,Schechter2008condmat}).

In this Letter we will focus on the low-concentration limit of these materials ($x = 0.045$) in zero field.  We have measured the ac magnetic susceptibility of this stoichiometry in the hope of reproducing the exotic physics that was observed previously.  We will show that, in fact, the susceptibility of this material behaves much more like that of a spin glass and that we are unable to reproduce the unusual antiglass phenomenology.

\begin{figure}
\begin{center}
\includegraphics[width=3.325in,keepaspectratio=true]{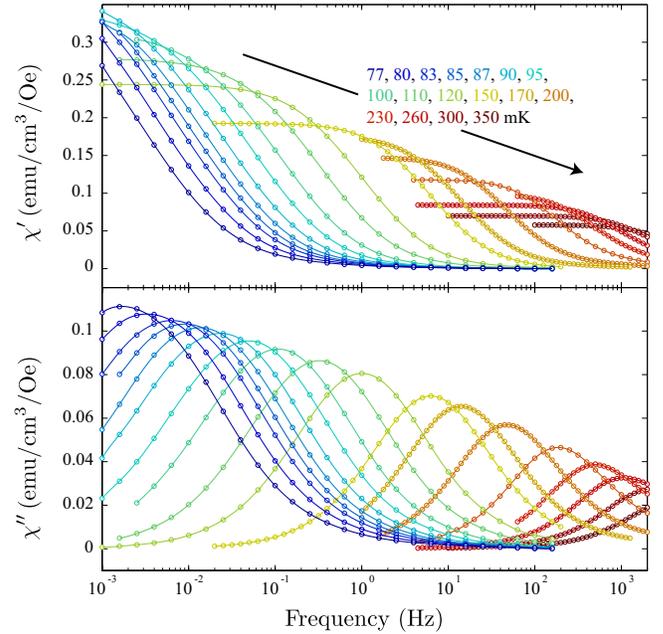}

\caption{\label{FrequencyScans}
AC susceptibility of the $x = 0.045$ sample showing in-phase $\chi'(f)$ and out-of-phase $\chi''(f)$ components.  The spectra were obtained at temperatures 77, 80, 83, 85, 87, 90, 95, 100, 110, 120, 140, 150, 170, 200, 230, 260, 300 and 350 mK from left (blue) to right (red).
}

\end{center}
\end{figure}


For this work, a magnetometer based on a dc superconducting quantum interference device (SQUID) was developed, chosen for its sensitivity and flat frequency response at very low frequencies.  The magnetometer consists of a 375-turn NbTi primary coil wrapped on a phenolic form surrounding a niobium, 2nd-order gradiometer.  The secondary forms one branch of a continuous superconducting loop or flux transformer which couples flux to a dc SQUID~\cite{EZSQUID}.  The flux transformer makes use of the Meissner effect meaning that our signal is not proportional to the frequency of excitation as it is in a standard inductive susceptometer.  This is crucial for attaining the low frequencies necessary to study such glassy dynamics.  An additional trim coil is used in parallel with the primary to remove misbalance of the gradiometer.

The SQUID is contained within a superconducting lead shield and is run in feedback mode.  The entire magnetometer apparatus is contained within a superconducting lead shield and the cryostat itself is surrounded by a lead-coated radiation shield and two $\mu$-metal shields.

Measurements were performed on a sample of LiHo$_x$Y$_{1-x}$F$_4$ with $x = 0.045$.  The sample was glued to a sapphire rod, the other end of which was heat sunk to the mixing chamber of a S.H.E. dilution refrigerator.  Previously, the specific heat of this same sample was measured and the sample was characterized with various X-ray scattering techniques as discussed in Ref.~\cite{Quilliam2007}.  For most measurements, the sample was cut to be needle-shaped (dimensions 0.57 mm $\times$ 0.77 mm $\times$ 7.7 mm) to reduce demagnetization effects with the long dimension along the $c$-axis (also the direction of the applied field).  The resulting demagnetization factor is $4\pi N = 0.493$ ~\cite{Aharoni1998}.  Some measurements were reproduced using the same sample cut to a different aspect ratio (0.57 mm $\times$ 0.77 mm $\times$ 3.31 mm) with demagnetization factor $4\pi N = 1.104$.  Matching the results for two differently shaped samples allowed us to calibrate our magnetometer and confidently determine the correct demagnetization correction.


Both detailed frequency scans at constant temperature and detailed temperature scans at constant frequency were obtained.  It was carefully checked that the sample was in equilibrium for all measurements by waiting several hours or more before measurement at a given temperature and also by taking multiple spectra to check for reproducibility.  The applied ac magnetic field was kept below 20 mOe at all times to ensure that there was no appreciable heating of the sample.  A range of fields around this value was tested and none were found to cause any significant difference in the spectra that might indicate a heating effect.  Measurements in the frequency range 0.001 Hz to 2 kHz are presented here.  The accessible frequency window of these experiments was limited at the high end by frequency dependent background signals and phase shifts, and the SQUID feedback electronics.  At the low end, it is fairly impractical to perform any measurements appreciably  below 1 mHz.


\begin{figure}

\begin{center}
\includegraphics[width=3.325in,keepaspectratio=true]{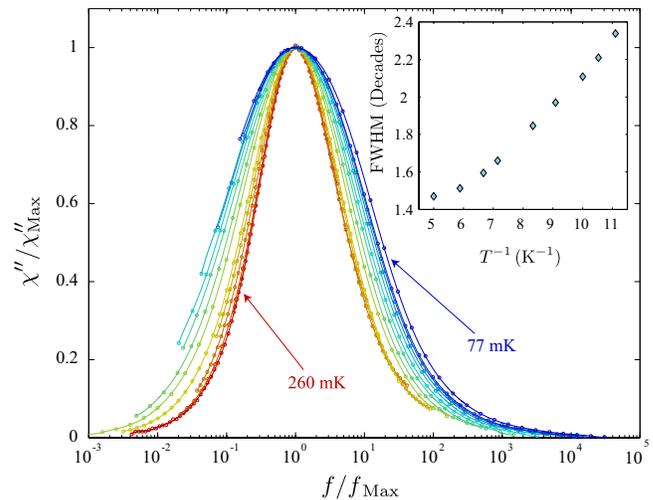}

\caption{\label{SuperimposedScans}
Frequency scans of $\chi''$ for various temperatures normalized by the peak height $\chi''_\mathrm{Max}$ on the vertical axis and by the peak frequency $f_\mathrm{Max}$ on the horizontal axis.  A clear broadening of the absorption spectra can be seen as the temperature is lowered from 260~mK to 77~mK.  The full width at half maximum (FWHM) in decades is shown in the inset as a function of inverse temperature.  The FWHM can only be obtained at a limited set of temperatures where the experimental frequency window envelops both sides of the curve.
}

\end{center}
\end{figure}

The measured frequency scans of the complex ac susceptibility are typical of glassy relaxation, showing broad peaks in $\chi''(\omega)$ and suppression of $\chi'(\omega)$ at higher frequencies as shown in Fig.~\ref{FrequencyScans}.  The characteristic time constants of relaxation, which may be parametrized by the peak frequencies in $\chi''$, $f_{\mathrm{\,Max}}$, are shifted to lower frequency with decreasing temperature as thermal fluctuations become weaker and are unable to excite the system over energy barriers.

Several studies of the $\chi(\omega)$ spectra of spin glasses have been performed previously including work on the LiHo$_x$Y$_{1-x}$F$_4$ series~\cite{Reich1990} and on other materials~\cite{Dekker1989,Huser1983, Alba1987}.  Despite being quite broad, the absorption spectra measured here are in fact narrow when compared to some other spin glass measurements~\cite{Dekker1989,Reich1990}.  Similar glassy relaxation has also been observed in the geometrically frustrated spin ice material Dy$_2$Ti$_2$O$_7$~\cite{Snyder2001,Matsuhira2001}.

Normalizing $\chi''(\omega)$ by the peak height $\chi''_{\mathrm{\,Max}}$ and dividing the frequency by the peak frequency $f_{\mathrm{\,Max}}$ superimposes the absorption spectra.  The result is a clear broadening of the spectra with decreasing temperature as seen in Fig.~\ref{SuperimposedScans}.  The full widths at half the maximum of the spectra (FWHM) are plotted in the inset of Fig.~\ref{SuperimposedScans}.  The FWHM appears to level off at $\sim 1.4$ decades at higher temperatures.   Though we have not shown frequency scans here, an 8\% sample also showed broadening of the absorption spectra with decreasing temperature.  This behavior is qualitatively similar with that of a higher concentration, $x=0.167$, material in this series~\cite{Reich1990}.

At the higher temperatures studied, the absorption spectra show low and high frequency limiting behavior of $\chi''\sim\omega^1$ and $\chi''\sim\omega^{-0.75}$ respectively.  This is not consistent with the Debye model with a single time constant where the limits are $\omega^1$ and $\omega^{-1}$.  As the temperature is reduced and the curves become broader, the tails of the spectrum become less steep.  At the lowest temperatures studied here, the high frequency limit can be seen to be as shallow as $\chi''\sim \omega^{-0.63}$.  These power laws constrain what functions might fit the data.  Several standard fitting functions including the Debye model with various distributions of relaxation times, the Davidson-Cole form~\cite{Davidson1951} and the Havriliak-Negami form~\cite{Alvarez1991} were considered but were not able to fit the full frequency range studied.

An important aspect of glassy systems is the critical slowing of dynamics or a divergence of the characteristic time constant of the system $\tau$.  The most robust way of determining $\tau$ would be to take the limit of $\chi''/\omega\chi'$ as $\omega\rightarrow 0$~\cite{Ogielski1985}, but such a limit is only achievable in a very small temperature window due to the broadness of the spectra and experimental time limitations.  Thus, in this work, we have chosen to parametrize the dynamics  of the system with $\tau_\mathrm{\,Max} = 1/2\pi f_\mathrm{\,Max}$.

\begin{figure}
\begin{center}
\includegraphics[width=3.325in,keepaspectratio=true]{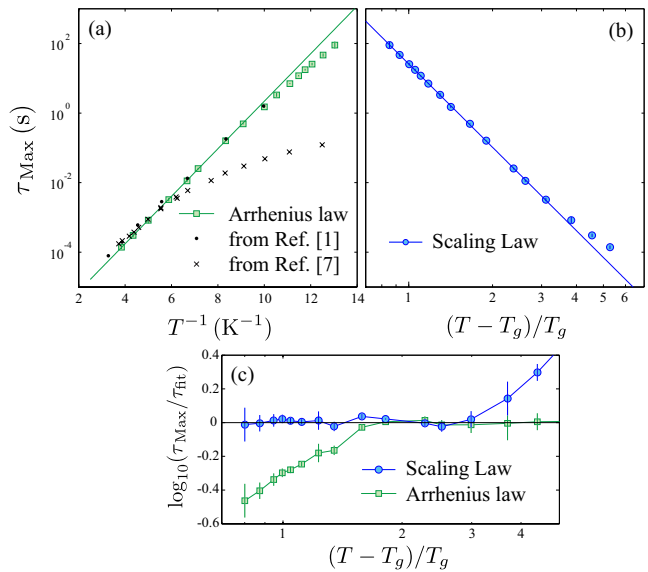}

\caption{\label{TauMaxFitting}
Different scenarios for scaling of the time constants obtained from the peak frequencies of the absorption spectra, $\tau_\mathrm{\,Max}(T)$.  (a) Data from this work fitted by an Arrhenius law at high $T$ (straight line).  Also shown are the time constants $\tau_\mathrm{\,Max}$ from Ref.~\cite{Reich1987} (dots) and Ref.~\cite{Ghosh2002} (crosses).  (b) Data from this work (circles) fitted by a dynamical scaling law at low $T$. (c) Residuals and estimated error bars for the Arrhenius and scaling law fits.
}

\end{center}
\end{figure}

At first glance, these peak positions appear to roughly follow an Arrhenius law, $\tau_{\mathrm{\,Max}}(T) = \tau_{0A} \exp\left(-E_A/k_B T \right)$, at least at higher temperatures, as shown in Fig.~\ref{TauMaxFitting}(a), giving $E_A = 1.57$ K and $\tau_{0A} = 0.32$~$\mu$s.  However there is some curvature thus the Arrhenius fit is only able to accommodate the higher temperature data points and significant deviation occurs at lower $T$.   A likely scenario at lower $T$ is a dynamical scaling law of the form $\tau_\mathrm{\,Max} = \tau_0 (T/T_g -1)^{-z\nu}$, that is predicted to apply to spin glasses~\cite{Ogielski1985}.  Such a fit, plotted in Fig.~\ref{TauMaxFitting}(b), is quite successful for the data below $\sim 200$ mK giving a transition temperature $T_g = 42.6$ mK $\pm 2$ mK and an exponent of $z\nu = 7.8 \pm 0.23$, very near the exponent 7.9 obtained in Monte Carlo simulations~\cite{Ogielski1985}.  Above 200 mK, as we move further from $T_g$ and out of the critical regime, the power law behavior breaks down and appears to give way to an Arrhenius law.

The overall intrinsic time scale of this material, $\tau_0 \simeq 20$~s, extracted from the power law fit, is extremely long (in Eu$_{0.4}$Sr$_{0.6}$S, for example, $\tau_0 \simeq 2\times10^{-7}$ s~\cite{Bontemps1984})  and explains why obtaining equilibrium data anywhere close to the transition temperature becomes completely impractical.  For example, measurements with a frequency of $\sim3\times10^{-5}$ Hz would be required to properly study even $T=1.5T_g$.  This difficulty would also apply to other measurements such as nonlinear susceptibility.  The near-Arrhenius law that is observed is likely a consequence of measuring at temperatures far above $T_g$.


\begin{figure}
\begin{center}
\includegraphics[width=3.325in,keepaspectratio=true]{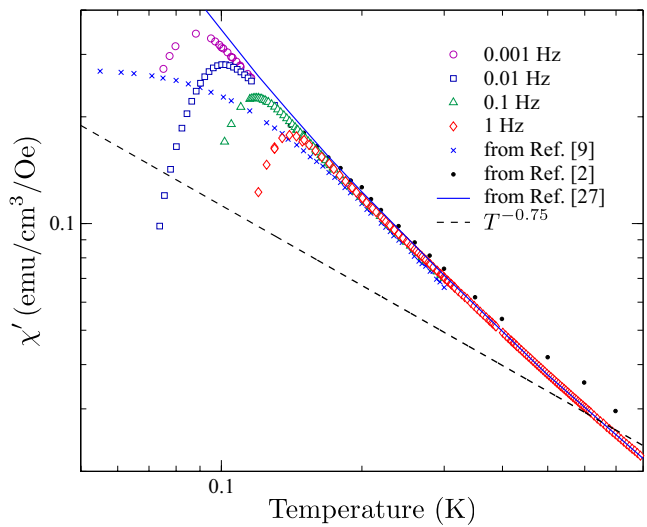}

\caption{\label{TempScans}
Temperature scans of $\chi'$ at four different frequencies of excitation.  We show for comparison experimental and theoretical results from other research groups: the DC limit of $\chi$ from Ref.~\cite{Reich1990} (as published), $\chi'$ taken with a linear field sweep from Ref.~\cite{Jonsson2007} (after correcting for demagnetization), Monte Carlo simulations~\cite{Biltmo2008} (arbitrary units, scaled to match at high $T$) and a $T^{-0.75}$ power law that was proposed in Ref.~\cite{Ghosh2003}.
}

\end{center}
\end{figure}

Detailed temperature scans at constant frequencies were also obtained and $\chi'(T)$ is shown in Fig.~\ref{TempScans}.  Predictably, one sees an increase in $\chi'$ with lower temperature until the frequency of relaxation of the sample becomes slower than the probe frequency.  At higher temperatures, where the sample's time constant is fast, the susceptibility measurements shown can be considered to be in the limit of static susceptibility. However, below approximately 90 mK, the sample's relaxation is so slow that even 1 mHz cannot be considered to be in the dc limit and there is a downturn in $\chi'$.  Nevertheless above 100 mK or so, we can effectively probe the static susceptibility of the material.


Comparison of these results with those obtained by Reich \emph{et al.}~\cite{Reich1987,Reich1990} shows a fair bit of agreement in certain respects.  In both cases the peak frequencies roughly follow an Arrhenius law and match \emph{quantitatively} as shown in Fig.~\ref{TauMaxFitting}(a).  The static limit of the magnetic susceptibility shows somewhat similar behavior as a function of temperature (see Fig.~\ref{TempScans}).  However, our results show that the widths of the absorption spectra do not narrow but rather broaden with lower temperature, a result more consistent with a spin glass.  In other words, we do not see ``antiglass'' physics reported in Ref.'s ~\cite{Ghosh2002,Ghosh2003}.


Papers published more recently~\cite{Ghosh2002,Ghosh2003} show results that differ strongly both from work by the same research group~\cite{Reich1987,Reich1990} and from our results presented here.  The shape and width of the absorption spectra are qualitatively different in Ref.~\cite{Ghosh2002} from the results in this work.  The peak frequencies also do not match with Ref.'s~\cite{Reich1987,Reich1990} and clearly do not follow an Arrhenius law, as shown in Fig~\ref{TauMaxFitting}(a).  Additionally, Ref.~\cite{Ghosh2003} states that the static susceptibility obeys a power law $T^{-\alpha}$ with an unusual $\alpha = 0.75$, a much shallower temperature dependence from that seen in this work and by Reich \emph{et al.}~\cite{Reich1987,Reich1990} (see Fig.~\ref{TempScans}).  Previous work by the authors of this paper~\cite{Quilliam2007}, measuring the specific heat of several stoichiometries in this series, found consistent, smooth features also in disagreement with Ref.'s~\cite{Reich1990,Ghosh2003} where unusual, sharp features were observed.

The ac susceptibility of LiHo$_{0.045}$Y$_{0.955}$F$_4$ was also measured recently by a third research group~\cite{Jonsson2007} and a third, distinct temperature dependence is observed.   At higher temperatures, where the response is fast, Ref.~\cite{Jonsson2007} matches well with our results here.  At lower temperatures, the discrepancy is likely a result of the magnetic field sweep rates used in Ref.~\cite{Jonsson2007} corresponding to frequencies of measurement that are too fast to be considered in the static limit and leading to an apparent reduction of $\chi$ (see Fig.~\ref{TempScans}).

We also compare our results to very recent Monte Carlo simulations of this material~\cite{Biltmo2008} and find that there is indeed very close agreement with the static limit of the susceptibility as a function of temperature (Fig.~\ref{TempScans}).  There is also qualitative agreement between the specific heat measured in Ref.~\cite{Quilliam2007} and that calculated in the simulations~\cite{Biltmo2008}.  These and other recent Monte Carlo simulations~\cite{Snider2005,Biltmo2007,Biltmo2008} find no divergence of the spin glass susceptibility suggesting that there is no finite temperature spin glass transition even if this system is viewed as a perfect Ising model.


However, classically, the mean-field theory of Aharony and Stephen~\cite{Aharony1981} would lead us to believe that there should be a spin glass transition all the way down to $x=0$.  It has been suggested by Ghosh \emph{et al.}~\cite{Ghosh2003} that off-diagonal terms inherent in the dipolar interaction can introduce quantum fluctuations, leading to a spin liquid state.  Recent theoretical work~\cite{Schechter2008condmat} analyzing a more detailed model incorporating random fields resulting from  the hyperfine interaction and off-diagonal dipolar coupling maintains that quantum effects are not strong enough to stabilize a spin liquid state and that the system should therefore undergo a spin glass transition.  They estimate a $T_g$ of roughly 35 mK, quite close to the 42.6 mK that we have determined in this work.  Additionally, Schechter and Stamp~\cite{Schechter2008condmat} predict a significant slowing of the dynamics as the concentration $x$ is lowered to 4.5\% as a result of the nuclear hyperfine coupling.

In conclusion, our measurements of LiHo$_x$Y$_{1-x}$F$_4$ have not shown the exotic antiglass physics that was observed previously~\cite{Reich1987,Reich1990,Ghosh2002,Ghosh2003}.  Instead, the absorption spectrum broadens with lower temperature, consistent with behavior expected of a spin glass. A dynamical scaling analysis using $\tau_\mathrm{\,Max}(T)$ obtained from the maxima in $\chi''$ implies the existence of a spin glass transition around $T_g = 43$~mK.  While the temperature dependence of the static susceptibility (Fig.~\ref{TempScans}) and specific heat~\cite{Quilliam2007} of this sample are similar to the Monte Carlo simulations~\cite{Biltmo2008}, dynamical scaling provides compelling evidence for a finite temperature spin glass transition, a conclusion which is supported by the theoretical work of Ref.~\cite{Schechter2008condmat}.   An extremely slow response of the system has also been observed and illustrates clearly the need for very low frequency measurements when studying dilute LiHo$_x$Y$_{1-x}$F$_4$ and other similar systems.  The ongoing debate~\cite{Biltmo2007, Jonsson2007,AnconnaTorres2008,Jonsson2008} on the existence of a spin glass transition at higher concentration LiHo$_{0.167}$Y$_{0.833}$F$_4$ might also be resolved through more careful attention to the diverging time scales involved.



\begin{acknowledgments}
We have benefited greatly from discussions with M.~J.~P. Gingras.  Thanks also to S.~W.~Kycia and A.~Gomez for discussions and help with sample characterization.  Funding for this research was provided by NSERC, CFI, MMO and Research Corporation grants.
\end{acknowledgments}


\begin{thebibliography}{10}


\bibitem{Reich1987}
D. H. Reich, T. F. Rosenbaum, and G Aeppli,
\newblock Phys. Rev. Lett. {\bf 59}, 1969 (1987).

\bibitem{Reich1990}
D.~H. Reich {\em et~al.},
\newblock Phys. Rev. B {\bf 42}, 4631 (1990).

\bibitem{Wu1991}
W. Wu, B. Ellman, T. F. Rosenbaum, G. Aeppli, and D. H. Reich,
\newblock Phys. Rev. Lett. {\bf 67}, 2076 (1991).

\bibitem{Wu1993}
W. Wu, D. Bitko, T. F. Rosenbaum, and G. Aeppli,
\newblock Phys. Rev. Lett. {\bf 71}, 1919 (1993).

\bibitem{Aharony1981}
A. Aharony and M. J. Stephen,
\newblock J. Phys. C {\bf 14}, 1665 (1981).

\bibitem{Ghosh2002}
S. Ghosh, R. Parthasarathy, T. F. Rosenbaum, and G. Aeppli,
\newblock Science {\bf 296}, 2195 (2002).

\bibitem{Ghosh2003}
S. Ghosh, T. F. Rosenbaum, G. Aeppli, and S. N. Coppersmith,
\newblock Nature {\bf 425}, 48 (2003).

\bibitem{Biltmo2007}
A. Biltmo and P. Henelius,
\newblock Phys. Rev. B {\bf 76}, 054423 (2007).

\bibitem{Jonsson2007}
P. E. J{\"o}nsson, R. Mathieu, W. Wernsdorfer, A. M. Tkachuck, and B. Barbara,
\newblock Phys. Rev. Lett. {\bf 98}, 256403 (2007).

\bibitem{AnconnaTorres2008}
C. Ancona-Torres, D. M. Silevitch, G. Aeppli, and T. F. Rosenbaum,
\newblock Phys. Rev. Lett. {\bf 101}, 057201 (2008).

\bibitem{Jonsson2008}
P. E. J{\"o}nsson {\em et~al.},
\newblock cond-mat/0803.1357 .

\bibitem{Schechter2005}
M.~Schechter and P.~C.~E. Stamp,
\newblock Phys. Rev. Lett. {\bf 95}, 267208 (2005).

\bibitem{Tabei2006}
S. M. A. Tabei, M. J. P. Gingras, Y. J. Kao, P. Stasiak, and J.-Y. Fortin,
\newblock Phys. Rev. Lett. {\bf 97}, 237203 (2006).

\bibitem{Schechter2008condmat}
M.~Schechter and P.~C.~E. Stamp,
\newblock cond-mat/0801.2889 .

\bibitem{EZSQUID}
SQUIDs and controller obtained from EZ-SQUID .

\bibitem{Quilliam2007}
J. A. Quilliam, C. G. A. Mugford, A. Gomez, S. W. Kycia, and J. B. Kycia,
\newblock Phys. Rev. Lett. {\bf 98}, 037203 (2007).

\bibitem{Aharoni1998}
A. Aharoni,
\newblock J. Phys. C {\bf 83}, 3432 (1998).

\bibitem{Dekker1989}
C. Dekker, A. F. M. Arts, H. W. de Wijn, A. J. van~Duyneveldt, and J. A. Mydosh,
\newblock Phys. Rev. B {\bf 40}, 11243 (1989).

\bibitem{Huser1983}
D. H\"{u}ser, L. E. Wenger, A. J. van~Duyneveldt, and J. A. Mydosh,
\newblock Phys. Rev. B {\bf 27}, 3100 (1983).


\bibitem{Alba1987}
M. Alba, J. Hammann, M. Ocio, P. Refregier, and H. Bouchiat,
\newblock J. Appl. Phys. {\bf 61}, 3683 (1987).

\bibitem{Snyder2001}
J. Snyder, J. S. Slusky, R. J. Cava, and P. Schiffer,
\newblock Nature {\bf 413}, 48 (2001).

\bibitem{Matsuhira2001}
K. Matsuhira, Y. Hinatsu, and T. Sakakibara,
\newblock J. Phys. Condens. Matter {\bf 13}, L737 (2001).

\bibitem{Davidson1951}
D. W. Davidson and R. H. Cole,
\newblock J. Chem. Phys. {\bf 19}, 1484 (1951).

\bibitem{Alvarez1991}
F. Alvarez, A. Alegr\'{i}a, and J. Colmenero,
\newblock Phys. Rev. B {\bf 44}, 7306 (1991).

\bibitem{Ogielski1985}
A. T. Ogielski,
\newblock Phys. Rev. B {\bf 32}, 7384 (1985).

\bibitem{Bontemps1984}
N. Bontemps, J. Rajchenbach, R. V. Chamberlin, and R. Orbach,
\newblock Phys. Rev. B {\bf 30}, 6514 (1984).

\bibitem{Biltmo2008}
A. Biltmo and P. Henelius,
cond-mat /0803.0851 .

\bibitem{Snider2005}
J.~Snider and C.~C. Yu,
\newblock Phys. Rev. B {\bf 72}, 214203 (2005).


\end{thebibliography}
\end{document}